\documentclass[aps,prb,twocolumn,amsmath,showpacs]{revtex4}

\usepackage{amssymb}
\usepackage{graphicx}
\usepackage{color}

\begin{document}

\title{Spatial variation of the laser fields and electron dynamics at a gas-solid interface}

\author{Georges Ra\c{s}eev\footnote{Corresponding author, E-mail: georges.raseev@ppm.u-psud.fr}}
\author{Eric Charron}

\affiliation{Laboratoire de Photophysique Mol\'{e}culaire du CNRS\footnote{Laboratoire de Photophysique Mol\'eculaire is associated to Universit\'e Paris-Sud.},\\ B\^{a}timent 210, Universit\'{e} Paris-Sud, 91405 Orsay Cedex, France}


\begin{abstract}
In the long wavelength domain, typically for wavelengths \mbox{$\lambda \geqslant 100$~\AA}, the laser fields are usually taken as independent of the spatial coordinate. However, at the gas-solid interface the electron density of the material and the incident laser fields vary sharply on a scale of few angstr\"oms. Instead of solving Maxwell equations, we present here a theoretical model, called \textit{Electromagnetic Fields from Electron Density} (EMFED), generating a continuous vector potential $\vec{A}\left(\vec{r},t\right)$ from phenomenological relations combining the unperturbed electron density of the material system, the material constants and the laws of optics. As an application of this model, we calculate in a time dependent approach the transition probability and the induced current density between the last bulk state below the Fermi energy and the first image state of a Cu(001) metallic surface. These observables are significantly modified by the spatial variation of the vector potential at the surface. The Coulomb gauge condition ($\vec{\nabla}\cdot\vec{A}=0$), fullfilled everywhere else, breaks down near the surface. The difference between the $\mathfrak{s}$- and $\mathfrak{p}$- polarizations of the laser field partially unravels this effect.
\end{abstract}

\maketitle

\section{Introduction}

Because of its importance for the photoelectric effect, many studies~\cite{makinson:49,adawi:64,mahan:70_b,endriz:73} have modeled the spatial variation of a laser electric field at a gas-solid (metal) interface in the long wavelength (LWL) domain (wavelengths $\lambda \geqslant 100$~\AA\, or, equivalently, energies \mbox{$\hbar\omega \leqslant$ 124 eV}). This electric field has been used to calculate the laser-matter interaction and the associated photoelectric probability. In 1975 Feibelman~\cite{feibelman:75} has evaluated the spatial dependence of the laser vector potential for a gas-solid Jellium interface solving the Maxwell equations using a nonlocal conductivity tensor. This model has been applied by Levinson~\textit{et~al.}~\cite{levinson:79} to photoemission and the results compared to experimental measurements.

More recently, Miller~\textit{et~al.}~\cite{miller:chiang:1997} have used the one step model of Mahan~\cite{mahan:70_b}, built on a Jellium potential displaced towards the vacuum to take into account a surface state, to study the direct, indirect and surface one photon photoemission of the Ag(111) surface. As in the work presented in this paper, this model uses a laser-matter operator in the velocity gauge written as a sum of a standard  $\vec{A}\cdot\vec{\nabla}$ and a surface $(\vec{\nabla}\cdot\vec{A})$ interaction term. Instead of an explicit calculation with a vector potential dependent on the spatial coordinate, the surface term has been fitted to the experimental spectrum to explain the unsymmetrical line shapes present in these data. Other models, derived from the work of Miller~\textit{et~al.}, have used the velocity gauge in the analysis of a two-photon photoemission experiment of metallic silver~\cite{pontius:petek:2005} near the surface plasmon resonance.

Using the same velocity form of the laser-matter interaction, Tergiman~\textit{et~al.}~\cite{tergiman:girardeau:97} have calculated the contribution to a photo-current of a multiphoton excitation by solving analytically the Schr\"odinger equation for the electrons. These authors have modified the Fermi-Dirac distribution describing the initial state of the electrons to include the contributions from electron-electron collisions. With this model, Tergiman~\textit{et~al.} were able to evaluate the linear and nonlinear multiphoton contributions to the photo-current as a function of the laser fluence.

In recent years many experimental studies in photoelectron emission have used the versatile two-photon photoemission (TPPE) technique~\cite{berthold:hofer:2004,kubo:petek:2005,kirchmann:wolf:2005,rohmer:aeschlimann:bauer:2006} to study the temporal evolution of the electron dynamics and of the associated observables on clean and covered surfaces. Today the majority of theoretical simulations of TPPE are performed using the density matrix theory in its Liouville-von Neumann form~\cite{wolf:1999,klamroth:saalfrank:hofer:2001,boger:weinelt:2002,pontius:petek:2005,boger:fauster:weinelt:2005,mii:ueba:2005}. The density matrix formulation has the advantage of permitting the inclusion of the elastic and inelastic electron-electron collisions which constitute efficient desexitation channels when the physical system is in interaction with the bath of the solid. Note that, except for the work of Pontius~\textit{et~al.}~\cite{pontius:petek:2005}, these density matrix formulations make use of the length gauge and the laser electric field is assumed to be independent of the spatial coordinate.

In this article we discuss the laser-matter interaction at a gas-solid interface in the long wavelength domain. The sharp rise of the electron density at the gas-solid interface, taking place on a sub-nanometric scale, affects the electromagnetic fields through a steep variation of the refractive index even if the particular atoms of the solid are not directly ``seen" by the electromagnetic fields. We try to answer the following question: what is the influence of the spatial dependence of the laser fields on the electron excitation at the interface and on the associated observables? This question may be of importance especially in the presence of adsorbates and nano structures. In a formulation where the vector potential $\vec{A}$ is a function of $\vec{r}$, there is an additional contribution to the laser-matter interaction coming from the operator $(\vec{\nabla}\cdot\vec{A})$ mentioned above. This contribution is known as a ``surface photoelectric effect'' (see e.g. Desjonqu\`eres and Spanjaard~\cite{book:desjonqueres_spanjaard}).

The above question is answered here in the framework of the Schr\"odinger equation, for a clean metallic surface in the context of a single excitation between a discretized state of the band of the solid and an image state. We have developed a phenomenological model, called \textit{Electromagnetic Fields from Electron Density} (EMFED), where the electromagnetic fields are explicitly function of the coordinate $z$ normal to the surface. A simplified version of this model, where the electron density used to generate the vector potential was calculated from a Jellium solid, has already been published~\cite{laser:matter:raseev:03, des:quant:bejan:03,emed:al:raseev:2005}. Here, the laser fields are calculated using macroscopic material constants, conductivity $\sigma$, dielectric function $\varepsilon$ and refractive index $\tilde{n}$, which depend on the chemical nature of the bulk material and on the laser wavelength. These material constants, taken from experimental measurements tabulated by Palik~\cite{book:palik_opt_cont,book:palik_opt_cont_II}, are considered to be local relative to the spatial coordinate $z$, \emph{i.e.} $\varepsilon(z,z')\simeq \varepsilon(z)$. To explicitly obtain the $z$-dependence of the laser field, the dielectric function $\varepsilon$ of the metal is related to the unperturbed electron density $\rho_e(z)$ of the metallic surface evaluated from the corresponding wave functions of the occupied states of the system. As in the Jellium calculations performed by Lang and Kohn~\cite{density:lang:kohn:70} using DFT (see also Jenning~\textit{et~al.}~\cite{electron:jenning:1988}), the calculated electron density is normalized with respect to its value in the bulk. The dielectric function $\varepsilon(z)$ is then connected to the vector potential $\vec{A}\left(\vec{r},t\right)$ through its wave vector $\vec{k}^{ph}$ \emph{via} the refractive index $\tilde{n}$.

Using the EMFED vector potential $\vec{A}\left(\vec{r},t\right)$, the Hamiltonian describing the interaction of the electrons of a metallic surface with the laser field is constructed. In the actual implementation of the model, the motion parallel and perpendicular to the surface are decoupled. Consequently, the wave function is expanded in terms of products of parallel and perpendicular to the surface basis functions. In the directions parallel to the surface we use a linear combination of atomic orbitals which fulfills the Bloch periodicity condition, and the set of functions perpendicular to the surface is, for simplicity, restricted to a single resonant contribution. The time dependent Schr\"odinger equation is projected on this basis, yielding a system of coupled first order ordinary differential equations (ODE) for the time variable. This ODE system of coupled equations is propagated for the duration of the laser pulse and the excitation probability and electron current density are calculated.

Specifically we have studied a clean Cu(001) surface where the bulk continuum near the Fermi level is discretized and we have considered a resonant excitation between this bulk Fermi and the first image states. The laser fluence used in the present simulation is in the range of the experiments by Velic~\textit{et~al.}~\cite{velic:wolf:98} for C$_6$H$_6$ on Cu(111) or Kirchmann~\textit{et~al.}~\cite{kirchmann:wolf:2005} for C$_6$F$_6$ on Cu(111), and is well within the perturbation regime.

\section{The laser field at the interface: the EMFED model}
\label{sec:EMFED_model}
\subsection{The vector potential of the laser field derived from the laws of optics}
\label{sec:A_pot}

\begin{figure}[htb!]
\centering
\includegraphics[width=8cm,clip]{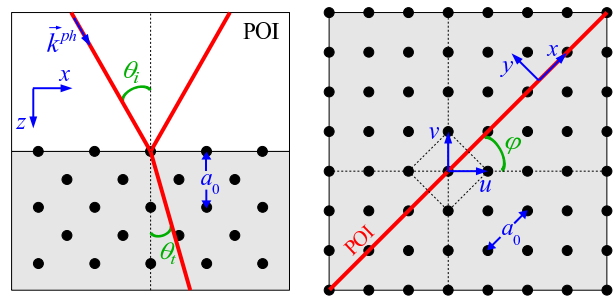}
\caption{(color online) Schematic picture of reflexion and refraction of a polarized laser beam incident on a metallic (001) surface of a face centered cubic (FCC) crystal. The FCC lattice parameter $a_0$ is shown and the shortest distance between two atoms on the surface corresponds to $a_0^u=a_0^v=a_0/\sqrt{2}$ in the $u$ and $v$ directions. The vector potentials $^{\mathfrak{s}}\!\vec{A}$ and $^{\mathfrak{p}}\!\vec{A}$, respectively for ${\mathfrak{s}}$- and ${\mathfrak{p}}$- polarizations, point in the directions perpendicular and parallel to the POI or $xOz$ (see text for details).}
\label{fig:Geometry}
\end{figure}

The geometry considered here is depicted in Figure~\ref{fig:Geometry}: the plane of incidence (POI) of the laser beam on the surface is defined as $(xOz)$, $z$ being the direction normal to the surface, pointing towards the metal. The three-dimensional space is therefore divided in two parts:  $z < 0$ corresponds to the gas phase and $z \geqslant 0$ corresponds to the solid. The angles $\theta_i$ and $\theta_t$ stand for the incident and transmitted angles. The angle between the POI and the crystallographic direction $\vec{u}$ of the surface is $\varphi$. The incoming electromagnetic wave, whose wave vector is denoted by $\vec{k}^{ph}$, is $\mathfrak{p}$- polarized when the vector potential $\vec{A}\left(\vec{r},t\right)$ (notation $^{\mathfrak{p}}\!\vec{A}$) is in the POI, and $\mathfrak{s}$- polarized when this vector (notation $^{\mathfrak{s}}\!\vec{A}$) is normal to the POI.

In the following, we label $(i)$, $(r)$ and $(t)$ the incident, reflected and transmitted components of the vector potential, taken as classical here. Taking into account the continuity of the fields at the interface (see e.g. Jackson~\cite{book:jackson}), one writes the normal projection (coordinate $z$) of the vector potential of a $\mathfrak{p}$- polarized monochromatic laser beam as
\begin{eqnarray}
\label{eq:A_z}
\left\{\begin{array}{ccll}
^{\mathfrak{p}}\!A_z(x,z,t) & = & \sin\theta_i \left[^{\mathfrak{p}}\!A^{(i)} + \,^{\mathfrak{p}}\!A^{(r)}\right] & (z < z_P)\\
^{\mathfrak{p}}\!A_z(x,z,t) & = & \sin\theta_t\; \tilde{\varepsilon}(\omega,z)
\;^{\mathfrak{p}}\!A^{(t)}  & (z \geq z_P),
\end{array}
\right.
\end{eqnarray}
and the tangent projection (transverse coordinate $x$) as
\begin{eqnarray}
\label{eq:A_x}
\left\{\begin{array}{ccll}
^{\mathfrak{p}}\!A_x(x,z,t) & = & \cos\theta_i \left[^{\mathfrak{p}}\!A^{(i)} - \,^{\mathfrak{p}}\!A^{(r)}\right] & (z < z_P)\\
^{\mathfrak{p}}\!A_x(x,z,t) & = & \cos\theta_t\; ^{\mathfrak{p}}\!A^{(t)} & (z\ge z_P),
\end{array}
\right.
\end{eqnarray}
where $z_P$ is the position of the image plane which can be different from the position $z=0$ of the geometrical surface (see next subsection). In Eq.~(\ref{eq:A_z}), one should notice the appearance of the complex relative local dielectric function $\tilde{\varepsilon}(\omega,z)$, function of the $z$ coordinate. 

For a $\mathfrak{s}$- polarized monochromatic laser beam, the vector potential is directed perpendicular to the POI, and therefore is parallel to the $y$ direction in our coordinate system, and 
\begin{eqnarray}
\label{eq:A_y}
\left\{\begin{array}{ccll}
^{\mathfrak{s}}\!A_y(x,z,t) & = & ^{\mathfrak{s}}\!A^{(i)} + \,^{\mathfrak{s}}\!A^{(r)} & (z < z_P)
\\
^{\mathfrak{s}}\!A_y(x,z,t) & = & ^{\mathfrak{s}}\!A^{(t)} & (z\ge z_P).
\end{array}
\right.
\end{eqnarray}
The incident, reflected and transmitted components of the vector potential used above are simply given by the standard classical expressions
\begin{equation}
\label{eq:A_irt}
\begin{array}{llrrl}
^{\mathfrak{pol}}\!A^{(i)}(x,z,t) & = &\! &\! \!A_{0} \!&
			\!e^{i\left(k^{ph}_x x\,+\,k^{ph}_z z\,-\,\omega t\right)} 
\\
^{\mathfrak{pol}}\!A^{(r)}(x,z,t) & = &\! ^{\mathfrak{pol}}\!R(\omega,z) 
		&\!\! A_{0}\! & \!e^{i\left(k^{ph}_x x\,-\,k^{ph}_z z\,-\,\omega t\right)} 
\\
^{\mathfrak{pol}}\!A^{(t)}(x,z,t) & = &\!  ^{\mathfrak{pol}}T(\omega,z) 
		&\!\! A_{0}\! & \!e^{i\left(k^{ph}_x x\,+\,k^{ph}_z z\,-\,\omega t\right)}
\end{array}
\end{equation}
where $\mathfrak{pol}$ corresponds to the $\mathfrak{p}$ or $\mathfrak{s}$ linear polarizations. The reflection $^{\mathfrak{pol}}\!R(\omega,z)$ and transmission $^{\mathfrak{pol}}T(\omega,z)$ coefficients are given by
\begin{subequations}
\label{eq:A_irt_coeff_p}
\begin{eqnarray}
^{\mathfrak{p}}\!R(\omega,z) &=&	
		 \frac{\tilde{n}(\omega,z)\;\cos\;\theta_i\;-\;\cos\;\theta_t}
		{\tilde{n}(\omega,z)\;\cos\;\theta_i\;+\;\cos\;\theta_t},
\\
^{\mathfrak{p}}T(\omega,z) &=&	\frac{2\cos\;\theta_i} 
		{\tilde{n}(\omega,z)\;\cos\;\theta_i\;+\;\cos\;\theta_t},
\end{eqnarray}
\end{subequations}
and
\begin{subequations}
\label{eq:A_irt_coeff_s}
\begin{eqnarray}
^{\mathfrak{s}}\!R(\omega,z) &=&	
		\frac{\cos\;\theta_i\;-\;\tilde{n}(\omega,z)\cos\;\theta_t}
		       {\cos\;\theta_i\;+\;\tilde{n}(\omega,z)\;\cos\;\theta_t},
\\
^{\mathfrak{s}}T(\omega,z) &=&	\frac{2\cos\;\theta_i} 
		{\cos\;\theta_i\;+\;\tilde{n}(\omega,z)\;\cos\;\theta_t},
\end{eqnarray}
\end{subequations}
where the complex refraction index $\tilde{n}(\omega,z)$, also function of $z$ coordinate, is obtained from the EMFED model as described in the next subsection. The attenuation of the field in the metal is simply taken into account through a complex wave vector 
\begin{equation}
\label{eq:k_ph}
k^{ph} = \tilde{n}\;\omega/c.
\end{equation}
In Eq.~(\ref{eq:A_irt}), the normalization factor $A_{0}$ defines the amplitude of the incident laser vector potential. It can be calculated from the fluence $F$ or the intensity $\mathfrak{I}_0$ of the laser pulse. If the
vector potential is described by a $\sin^2$ pulse shape of full width at half maximum (FWHM) $\tau$, one gets
\begin{eqnarray}
\label{eq:A_0}
A_0(\mathrm{a.u.}) & = & \frac{1.66\;10^{-3}}{i\;\omega(\mathrm{a.u.})}\;\sqrt{\frac{F(\mathrm{J/m^2})}{\tau(\mathrm{fs})}}.
\end{eqnarray}
The numerical relation between the fluence $F$ and the laser intensity $\mathfrak{I}_0$ for this pulse shape is 
\begin{displaymath}
F(\mathrm{J/m^2})=0.9708\;\mathfrak{I}_0(\mathrm{W/m^2})\;\tau(\mathrm{s}).
\end{displaymath}
The angle of refraction $\theta_t$ is, in this phenomenological model, a complex function of the incident laser angular frequency $\omega$ and of the coordinate $z$. It is calculated from the law of Snell-Descartes
\begin{equation}
\label{eq:theta_t}
\sin\;\theta_t = \frac{\sin\;\theta_i}{\tilde{n}(\omega,z)}\;.
\end{equation}
%
\subsection{Material constants from the conductivity and the electron density}

The equations~(\ref{eq:A_z}-\ref{eq:A_y}) are standard expressions of continuity at an abrupt interface. Here the material constants, as for example the dielectric function $\tilde{\varepsilon}$ or the refractive index $\tilde{n}$, are functions of the coordinate $z$. This dependence also induces a dependence relative to $z$ of the associated wave number $k^{ph}$ (Eq.~(\ref{eq:k_ph})) and of the refraction angle $\theta_t$ (Eq.~(\ref{eq:theta_t})).

Following Drude's model~\cite{Drude}, at optical and higher frequencies, both bound and conduction electrons of a metal contribute to the dielectric function through the relation
\begin{equation}
\label{eq:Esigma}
\tilde{\varepsilon}(\omega)=\varepsilon_b(\omega) + i\;\frac{\tilde{\sigma}(\omega)}{\omega\,\varepsilon_0}\,,
\end{equation}
where $\varepsilon_b(\omega)$ is the relative permittivity due to the bound electrons, $\varepsilon_0$ denotes the vacuum permittivity (only non-magnetic materials are considered here with a relative permeability $\mu=1$) and $\tilde{\sigma}(\omega)$ is the complex conductivity~(see e.g. Jackson\cite{book:jackson}, chapter 7). The Drude model also gives the following analytical expression for the conductivity
\begin{equation}
\label{eq:conductivity}
\tilde{\sigma}(\omega) = \frac{e^2}{m^{*}\, (\gamma_0-i\;\omega)}\;\rho^s\,,
\end{equation}
where $e$ is the charge of the electron, $m^*$ its effective mass, $\rho^s$ the electron density of the bulk and $\gamma_0$ the electron friction coefficient.

The above equations relate $\tilde{\varepsilon}(\omega)$ and $\rho^s$ and can be used as a template to model in a simple way the spatial dependence of the dielectric function at the surface
\begin{equation}
\label{eq:dielectric_analytique}
\tilde{\varepsilon}(\omega,z) = 1 + \left(\tilde{\varepsilon}_s(\omega) - 1\right)\,\rho_e(z),
\end{equation}
where $\tilde{\varepsilon_s}(\omega)$ is the complex relative dielectric function in the solid and $\rho_e(z)$ is the relative electron density normalized with respect to the bulk density $\rho^s$. Equation~(\ref{eq:dielectric_analytique}) has the advantage of imposing a continuous variation of $\tilde{\varepsilon}(\omega,z)$ through the interface with a linear dependence on $\rho_e(z)$. In addition this expression shows the correct asymptotic limits
\begin{displaymath}
\left\{
\begin{array}{cccl}
\tilde{\varepsilon}(\omega,-\infty) & = & 1 &\\
\tilde{\varepsilon}(\omega,+\infty) & = & \tilde{\varepsilon}_s(\omega) & .
\end{array}
\right.
\end{displaymath}

Equation (\ref{eq:dielectric_analytique}) is the heart of the EMFED model which uses both the unperturbed relative electron density of the material system $\rho_e(z)$ under study and the tabulated values~\cite{book:palik_opt_cont,book:palik_opt_cont_II} of the associated complex refractive index of the bulk ($\tilde{n}_s^2(\omega)=\tilde{\varepsilon}_s(\omega)$).

\begin{figure}[htb!]
\centering
\includegraphics[width=8cm,clip]{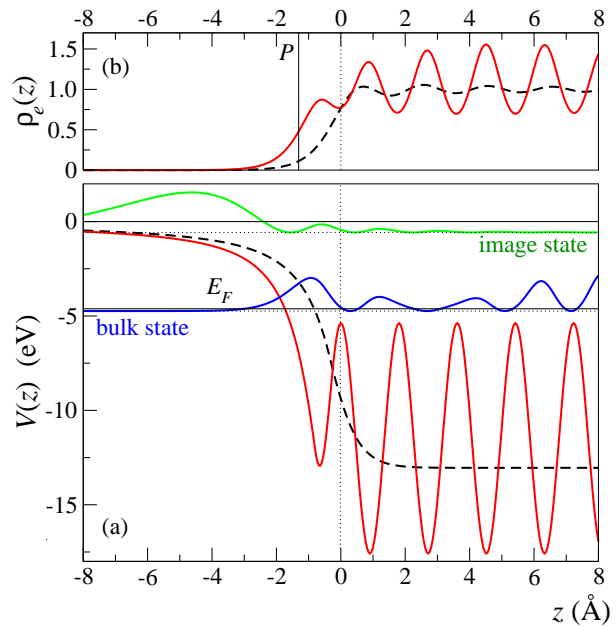}
\caption{(color online)  Lower graph (a): The electron-matter interaction potentials for Cu(001) using the Chulkov~\textit{et~al.} (red solid line) or Jellium with $r_s$=2.67~a.u. (black dashed line) models. The probability densities $\left|\eta_I(z)\right|^2$ associated with the bulk Fermi and first image states accommodated by the Chulkov potential are also shown at their energy (- 0.12~eV from $E_F$; -0.574~eV from the vacuum) in blue and green respectively. Upper graph (b): the total electron densities calculated using Eq.~(\ref{eq:electron_density}) with the same colors and line types. The vertical line at $z_P=-1.31\,$\AA\, corresponds to the position of the image plane $P$ at $\rho_e(z_P)=0.5$ where the continuity of the vector potential is enforced.}
\label{fig:total_density_cu_001}
\end{figure}

Using the density functional theory (DFT), the spatial variation of the electron density $\rho_e(z)$ at a gas-Jellium interface has been calculated by Lang and Kohn~\cite{density:lang:kohn:70} for several electrons densities in the bulk $\rho^s$, or Wigner-Seitz radii $r_s=(3/(4\pi\rho^s))^{1/3}$. A calculation of this electron density from the occupied states of the material system, using a smooth model Jellium potential at the interface due to Jennings~\textit{et~al.}\cite{electron:jenning:1988}, gives similar results. In the present implementation of the EMFED model to a metallic surface, we calculate this electron density from a more accurate model potential perpendicular to the surface parametrized by Chulkov~\textit{et~al.}~\cite{des:quant:chulkov:99}. The Jellium and Chulkov~\textit{et~al.} potentials are displayed as red solid and black dashed lines in the lower part of Figure~\ref{fig:total_density_cu_001}. For a given potential, the total relative electron density $\rho_e(z)$ is expressed as a discretized sum
\begin{eqnarray}
\label{eq:electron_density}\displaystyle
\rho_e(z) &\varpropto&\sum_{I=1,n_F} |\eta_I(z)|^2
\end{eqnarray}
running over all the occupied states up to the Fermi level. $\eta_I(z)$ denote here the wavefunctions associated with the eigenstates of the one-dimensional time-independent Schr\"odinger equation in the $z$ direction. This relative electron density $\rho_e(z)$ is normalized with respect to the mean average value of the density $\rho^s$ deep in the bulk. The sum is discretized because we have adopted a discrete variable representation (DVR) approach~\cite{dvr:bacic:light:1986,dvr:colbert:miller:1992} for the description of the continuum band structure of the metal. The electron densities calculated with the potential of Chulkov~\textit{et~al.} and with the Jellium potential are displayed as red solid and black dashed lines in the upper part of Figure~\ref{fig:total_density_cu_001}. For the Cu(001) surface studied here, the electrons at the Fermi level belong to the occupied band, and, because of the use of the DVR, the band is discretized. The last discrete state below the Fermi energy, hereafter called bulk Fermi state, and the first  image state ($n=1$) are displayed in the lower part of Figure~\ref{fig:total_density_cu_001}. These states participate in the photoexcitation considered here. Note that in the present case these states and the total electron density extend over several \"angstroms in the vacuum. Note also the difference in the amplitude of the oscillations of the densities and potentials between the Chulkov~\textit{et~al.} and Jellium models. This difference is due to the explicit inclusion of the atomic structure in the first but not in the second model.

The refraction index is obtained from the standand relation to the dielectric function 
\begin{equation}
\label{eq:refraction_index}
\tilde{n}(\omega,z)=\sqrt{\varepsilon(\omega,z)}\,.
\end{equation}
Using the electron density~(\ref{eq:electron_density}), the analytic form of the dielectric function~(\ref{eq:dielectric_analytique}), and the associated refraction index
one can calculate, from the expressions of subsection~\ref{sec:A_pot}, the vector potential $\vec{A}\left(\vec{r},t\right)$ of the laser field at the interface. This simple procedure
allows us to introduce the spatial variation of the vector potential at the gas-solid interface.

The EMFED model explained above can tentatively be understood from the point of view of the solutions of the Maxwell equations. Far in the vacuum the asymptotic condition corresponds to the incident plane wave $ ^{\mathfrak{pol}}\!A^{(i)}$ of Eq.~(\ref{eq:A_irt}). The propagation of the Maxwell equations towards the interface smoothly accumulates the presence of the electron density in the vector potential. At the end of the propagation far in the bulk the vector potential corresponds to the transmitted component $^{\mathfrak{pol}}\!A^{(t)}$. In the EMFED model, the reflected component contributes near the surface, therefore allowing for the continuity of the electromagnetic fields at the interface. This result is therefore qualitatively similar to a model of Feibelman~\cite{feibelman:75_l} who solved numerically Maxwell equations for a vacuum/Jellium system.

\begin{figure}[htb!]
\centering
\includegraphics[width=8cm,clip]{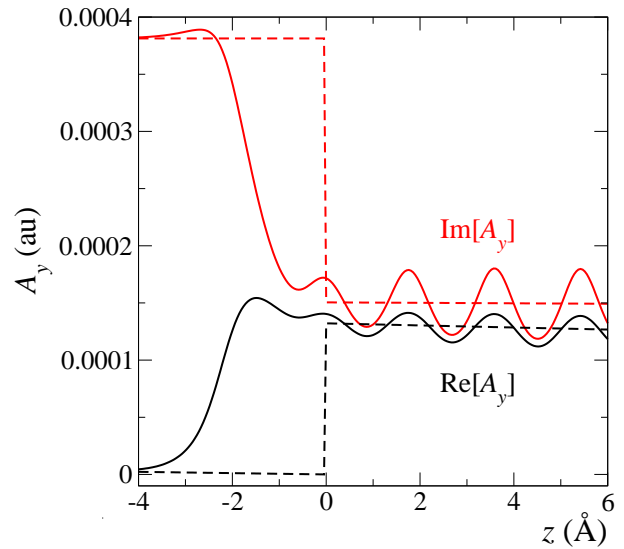}
\caption{(color online) Real and imaginary components $y$ of the vector potential $\vec{A}$ as a function of the coordinate $z$ for a $\mathfrak{s}$- polarized laser beam. Following Eq. (\ref{eq:A_irt}), the $x$ dependence in Eq. (\ref{eq:fields}) corresponds to a plane wave and therefore the dependence of the vector potential relative to this variable is not shown in the present figure. The photon energy and the laser fluence are 4.169~eV and 10~$\mu$J/cm$^2$ respectively. The real and imaginary refraction indexes of the bulk metallic Cu, $n$=1.386 and $\kappa$=1.687 at 4.2~eV, are taken from the tables of Palik~\cite{book:palik_opt_cont,book:palik_opt_cont_II}. The black Re$[\vec{A}]$ and red Im$[\vec{A}]$ solid lines are the results obtained with the EMFED model, while the dashed lines correspond to the incident and transmitted fields in a simple approach from optics with constant vector potentials (quotted as gas/solid in the figures below).}
\label{fig:cu_001_As_y}
\end{figure}

For the $\mathfrak{s}$- polarization, the vector potential presents a single $y$ component displayed in Figure~\ref{fig:cu_001_As_y}. It shows a smooth variation everywhere and particularly near the image plane of the Cu(001) system at \mbox{$z_P \simeq -1.31\,$\AA} corresponding to the electron density $\rho_e(z_P)=0.5$ where, following Eqs. (\ref{eq:A_z}),  (\ref{eq:A_x}) and  (\ref{eq:A_y}), the continuity of the vector potential is enforced. This image plane can be shifted from $\rho_e(z_P)=0.2$ to $\rho_e(z_P)=0.8$ with nearly no consequence on the vector potential itself. Similar smooth behavior is obtained in the calculation of Feibelman~\cite{feibelman:75_l}. The dashed lines in Figure~\ref{fig:cu_001_As_y} correspond to a calculation with a constant vector potential in each medium where only the incident and transmitted components are taken into account.

Various derivations and models have also used the material constants in the two media together with the continuity at the interface to obtain the corresponding properties of the laser-matter interaction. For example, Feibelman~\cite{feibelman:1989} obtained in this way the surface plasmon frequency and Liebsch and Schaich~\cite{liebsch:schaich:1995} explained the silver anomalous dispersion of the surface plasmon frequency.

Because we are not solving explicitly the Maxwell equations for the electromagnetic fields, the problem of the gauge used here has not been touched up to now. How behaves the external electromagnetic field in the two media (vacuum and bulk) and at the interface from the point of view of the gauge of the electromagnetic wave? The vector potential can be decomposed in perpendicular and parallel components with respect to the surface plane as $\vec{A}=\vec{A}_{\perp}+\vec{A}_{\parallel}$. A simple look at Figure~\ref{fig:Geometry} gives
\begin{equation}
\label{eq:fieldp}
\left\{
\begin{array}{cccl}
^\mathfrak{p}\!A_{\perp}     & = & ^\mathfrak{p}\!A_z(x,z)\\
^\mathfrak{p}\!A_{\parallel} & = & ^\mathfrak{p}\!A_x(x,z)
\end{array}
\right.
\end{equation}
for the $\mathfrak{p}$- polarization, and
\begin{equation}
\label{eq:fields}
\left\{
\begin{array}{cccl}
^\mathfrak{s}\!A_{\perp}     & = & 0\\
^\mathfrak{s}\!A_{\parallel} & = & ^\mathfrak{s}\!A_y(x,z)
\end{array}
\right.
\end{equation}
for the $\mathfrak{s}$- polarization. In this last case of $\mathfrak{s}$- polarization, it is already clear that \mbox{$\vec{\nabla} \cdot\, ^\mathfrak{s}\!\!\vec{A} = 0$} everywhere, and the Coulomb gauge condition is therefore verified in our model for this specific polarization. The case of $\mathfrak{p}$- polarization is clearly different since the fast variation of $^\mathfrak{p}\!A_{\perp}$ with $z$ in the vicinity of the surface implies that \mbox{$\vec{\nabla} \cdot\, ^\mathfrak{p}\!\!\vec{A} \neq 0$} when approaching the metal. As a consequence, the electromagnetic fields given by the EMFED model do not verify the Coulomb gauge for the $\mathfrak{p}$- polarization.
This discussion also applies to the Schr\"odinger equation, and the laser-matter interaction term in the Hamiltonian includes a non-Coulomb gauge term proportional to $(\vec{\nabla}\cdot\vec{A})$.

\section{A time dependent model for the electronic dynamics at the interface}
\label{sec:time_dependent_model}

\subsection{The Hamiltonian}
\label{sec:hamiltonian}

The total electronic Hamiltonian including the laser-matter interaction is similar to the Hamiltonian for the unperturbed system except that the electron momentum $\vec{p}$ is replaced by ($\vec{p}+e\vec{A}$), where $e$ and $\vec{A}(\vec{r},t)$ are the charge of the electron and the vector potential of the laser field. The total Hamiltonian reads
\begin{eqnarray}
\label{eq:Hamiltonian}
\hat{H}(\vec{r},t) & = & \hat{H}^0(\vec{r}) + \hat{H}^L(\vec{r},t) + \hat{H}^Q(\vec{r},t),
\end{eqnarray}
where the zero order Hamiltonian is 
\begin{equation}
\hat{H}^0(\vec{r}) =  \frac{\vec{p}\,^2}{2\;m} + V(\vec{r}).
\end{equation}

The laser-matter interaction terms of the Hamiltonian can be decomposed as
\begin{subequations}
\begin{eqnarray}
\label{eq:H_L}
\hat{H}^L(\vec{r},t) &=&\frac{\hbar e}{2 i m}\bigl[2\;
            \vec{A}\cdot \vec{\nabla} + (\vec{\nabla}\cdot\vec{A})\bigr]\\
\label{H_Q}
\hat{H}^Q(\vec{r},t) &=& \frac{e^2}{2m}[\vec{A}\cdot\vec{A}],
\end{eqnarray}
\end{subequations}
where the parentheses in the first equation above mean that the operator $\vec{\nabla}$ acts on the vector potential $\vec{A}(\vec{r},t)$ only. The superscripts $L$ and $Q$ stand respectively for the linear and quadratic terms in $\vec{A}$. 

\begin{table}[h!]
\caption{\label{tab:A_p_operator}Decomposition of the laser-matter interaction Hamiltonian for a $\mathfrak{p}$- and $\mathfrak{s}$- polarized vector potential, respectively $^\mathfrak{p}{\vec{A}(\vec{r},t)}$ and  $^\mathfrak{s}{\vec{A}(\vec{r},t)}$.}
\begin{ruledtabular}
\begin{tabular}{ccccc}
 & $\hat{H}_{\parallel}^L$ & $\hat{H}_{\perp}^L$ & $\hat{H}_{\parallel}^Q$ & $\hat{H}_{\perp}^Q$ \\
\hline
$\mathfrak{p}$ & $2\;A_x\;\frac{\partial}{\partial x} + \left(\frac{\partial A_x}{\partial x}\right)$ & $2\;A_z\;\frac{\partial}{\partial z} + \left(\frac{\partial A_z}{\partial z}\right)$ & $A_x^2$ & $A_z^2$\\
$\mathfrak{s}$ & $2\;A_y\;\frac{\partial}{\partial y}$ & ----- & $A_y^2$ & -----\\
\end{tabular}
\end{ruledtabular}
\end{table}

The analysis given in the preceding section has revealled a non-zero divergence of the vector potential $(\vec{\nabla}\cdot\vec{A})$ in the case of $\mathfrak{p}$- polarization. This non-Coulomb gauge contribution develops in two terms, one perpendicular \mbox{$(\partial A_z/\partial z)$}, and the other parallel $(\partial A_x/\partial x)$ to the surface (see Table \ref{tab:A_p_operator}). These two terms, called {\it surface terms}, give rise to the so called surface photoelectric effect (see e.g. Desjonqu\`eres and Spanjaard~\cite{book:desjonqueres_spanjaard}). For $\mathfrak{s}$- polarization, these two terms do not appear, but Table~\ref{tab:A_p_operator} shows that two standard contributions remain in the interaction Hamiltonian: $2\;A_y\partial /\partial y$ and $A_y^2$.

Table \ref{tab:A_p_operator} also shows that a characterization of the laser field can be performed by calculating the difference between the observables measured for $\mathfrak{p}$ and $\mathfrak{s}$- polarizations. For equivalent directions $x$ and $y$ parallel to the surface, the contributions from $A_x^2+2\;A_x\partial /\partial x$ of $\mathfrak{p}$ and $A_y^2+2\;A_y\partial /\partial y$ of $\mathfrak{s}$- polarizations are equivalent and eliminated from the interaction. Unfortunately  $A_z^2+2\;A_z\partial /\partial z$ is not eliminated and therefore one cannot measure directly the influence of the surface terms $(\vec{\nabla}\cdot \vec{A})=(\partial A_x/\partial x) + (\partial A_z/\partial z)$.

\subsection{The time dependent Schr\"odinger equation}
\label{subsec:time_dependent_model}

The time-dependent Schr\"odinger equation describing the interaction of the laser field with the electrons at the gas-solid interface reads
\begin{equation}
\label{eq:TDSE}
i\hbar\;\frac{\partial}{\partial t}\,\Psi(\vec{r},t) = \hat{H}(\vec{r},t)\,\Psi(\vec{r},t),
\end{equation}
where $\Psi(\vec{r},t)$ denotes the electronic time-dependent wavefunction. This wave function is expanded on a series of time independent basis functions $\Phi_I (\vec{r})$
\begin{eqnarray}
\label{eq:expansion}
\Psi(\vec{r},t) &=& \sum_I \Phi_I (\vec{r})C_I(t) = \mathbf{\Phi} \mathbf{C},
\end{eqnarray}
where $C_I(t)$ are the time dependent expansion coefficients. The second expression corresponds to a matrix notation.

Figure~\ref{fig:Geometry} shows that the presence of the two media breaks the periodicity in the direction $z$ normal to the surface. On the other hand, and since there is no electron density in the vacuum, we will treat the system as a periodic system in the parallel direction even in the vaccum. In the present version of the model we write the basis wave function $\Phi_I (\vec{r})$ of Eq.~(\ref{eq:expansion}) as a product
\begin{equation}
\label{eq:basis_functions}
\Phi_I(\vec{r}) =  \eta_I(z) \; f_I(\vec{r}_{\parallel})\,,
\end{equation}
where $\eta_I(z)$, already defined with Eq.~(\ref{eq:electron_density}), is a wavefunction associated with the eigenstates of the one-dimensional time-independent Schr\"odinger equation along the perpendicular direction $z$. Along the surface plane direction
\mbox{$\vec{r}_{\parallel}=(x,y)$}, the basis functions $f_I(\vec{r}_{\parallel})$ are written as
\begin{equation}
\label{eq:f_I}
f_I(\vec{r}_{\parallel}) = \sum_{j}\;\; e^{i\;\vec{k}_{\parallel}^I\cdot\vec{R}_{\parallel}^j}\;\;
                           \chi_{I}(\vec{r}_{\parallel}-\vec{R}_{\parallel}^j)\,,
\end{equation}
where $\vec{k}_{\parallel}^I$ denotes the parallel electron wave vector. In this expression, $\chi_{I}$ denotes a function defined in the elementary cell. The vectors $\vec{R}_{\parallel}^j$ allow for the translation of this localized wavefunction in the parallel direction. This generates a two-dimensional periodic function in the surface plane which fulfills the Bloch periodicity condition (see for instance Ziman~\cite{book:ziman}). This approach is similar to the method of linear combinations of atomic orbitals (LCAO) for a solid.

Inserting the expansion~(\ref{eq:expansion}) in the time-dependent Schr\"oringer equation~(\ref{eq:TDSE}), and projecting it on the time independent basis functions $\Phi_I(\vec{r})$ yields a coupled system of first order ordinary differential equations (ODE) for the expansion coefficients $C_I(t)$
\begin{equation}
\label{eq:coupled_equations}
i\hbar\frac{\partial C_I}{\partial t}  =  \sum_{J}\; \langle\Phi_I|\hat{H}(\vec{r},t)|\Phi_J\rangle\; C_J(t)\,.
\end{equation}
The diagonal terms can be written as
\begin{equation}
\label{eq:diagonal}
\langle\Phi_I|\hat{H}(\vec{r},t)|\Phi_I\rangle = E_I^{\perp} + E_I^{\parallel}
	  	+ \langle\Phi_I|\hat{H}^L+\hat{H}^Q|\Phi_I\rangle\,,
\end{equation}
where $E_I^{\perp}$ and $E_I^{\parallel}$ denote the energies associated with the wavefunctions $\eta_I(z)$ and  $f_I(\vec{r}_{\parallel})$ respectively.

For low laser intensities, the off-diagonal laser-matter interaction matrix element \mbox{$\langle\Phi_I|\hat{H}^L+\hat{H}^Q|\Phi_J\rangle$} can be approximated by its linear component \mbox{$\langle\Phi_I|\hat{H}^L|\Phi_J\rangle$}. Here this term is an integral over the entire space. On the other hand, the current density, Eq.~(\ref{eq:courant_I_int}) below, also contains this integral but taken over a restricted domain in the perpendicular direction, namely from $-\infty$ to $z$. One can write this integral, to be used in these two expressions, as
\begin{equation}
\label{eq:interaction}
\langle\Phi_I|\hat{H}^L|\Phi_J\rangle_z = \frac{\hbar\,e}{2\,i\,m}\;\mathcal{J}_{IJ}(z)\;e^{-i\,\omega\,t}\,,
\end{equation}
where the calculation of $\mathcal{J}_{IJ}(z)$ is detailed in Appendix~\ref{annex:A}.

Using the model explained above, two observables are calculated: the excitation probability and the electron current density. The population of the different eigenstates and therefore the excitation probability are obtained from the overlap between the states $\Phi_I(\vec{r})$ and the final wave packet $\Psi(\vec{r},t_f)$ at the end of the pulse ($t=t_f$)
\begin{equation}
\label{eq:section_efficace_I}
P_I(t_f) = \left|\;\langle\;\Phi_I(\vec{r})\;\left|\;\Psi(\vec{r},t_f)\;\right.\rangle\;\right|^2 = \left|C_I(t_f)\right|^2.
\end{equation}

The expression for the electron current density, including the linear momentum of the electron $\vec{p}$ and of the laser field $e\vec{A}$, can be written using the kinetic energy operator $\hat{T}=(\vec{p}+e\,\vec{A})^2/2\;m$ (see e.g. Messiah~\cite{book:messiah}) as
\begin{eqnarray}
\label{eq:courant_I}
I(z,t) & = & \frac{1}{i\hbar} \int_{-\infty}^z\!\!\!\!dz'\int_{S_{cell}}\!\!\!\!\!\!\!\!dS\;
\left[\Psi^{\dagger}\hat{T}\Psi-(\hat{T}\Psi)^{\dagger}\Psi\right],
\end{eqnarray}
where $S_{cell}$ denotes the surface of the elementary cell along the parallel direction. This total current can be split in two parts
\begin{eqnarray}
\label{eq:courant_I_0_int}
I(z,t)    & = & I_0(z,t)\;+\;I_{int}(z,t)\,,
\end{eqnarray}
with
\begin{subequations}
\begin{equation}
\label{eq:courant_I_0}
I_0 = \frac{\hbar}{2im}\mathbf{C}^{\dagger}\!\!\left[
\int_{-\infty}^z\!\!\!\!\!\!dz'\!\!\int_{S_{cell}}\!\!\!\!\!\!\!\!dS\;
\left(\mathbf{\Phi}^{\dagger}\nabla^2\mathbf{\Phi}-\nabla^2\mathbf{\Phi}^{\dagger}\mathbf{\Phi}
\right)\right]\!\!\mathbf{C}
\end{equation}
and
\begin{equation}
\label{eq:courant_I_int}
I_{int} = -\frac{e}{2m} \mathbf{C}^{\dagger}\left[
e^{-i\omega t}\,\mathbf{J}_{int}(z)-e^{i\omega t}\,\mathbf{J}_{int}^{\dagger}(z)
\right]\mathbf{C}\,.
\end{equation}
\end{subequations}
As in Eq.~(\ref{eq:expansion}), in the previous equations $\mathbf{\Phi}$ is the line vector of the basis functions
and $\mathbf{C}$ is the column vector of the coefficients. The term $\mathbf{J}_{int}(z)$ is a matrix whose elements are the integrals $\mathcal{J}_{IJ}(z)$ given in Eqs.~(\ref{eq:J_int_p}) and (\ref{eq:J_int_s}). In the calculation of $I_{int}$, the quadratic terms in $\vec{A}$ have been neglected.

\section{The physical model and the excitation dynamics}
\label{sec:physmod}

\subsection{Details of the physical model}
\label{subsec:physmod}

In the present model calculation we concentrate on the influence of the spatial dependence of the laser field in the long wavelength domain (section \ref{sec:A_pot}) on the observables associated with a photoexcitation process. 

The model representing the material system is restricted to its simplest meaningfull form which is summarized below. First, the clean Cu(001) surface has been selected since it corresponds to a simple surface structure. The vectors $\vec{u}$ and $\vec{v}$, used to generate the (001) surface displayed in Figure~\ref{fig:Geometry}, are not the primitive vectors of the FCC solid and the associated potential $V(u,v,z)$ is not separable. For example, in the plane of incidence $(xOz)$ there is a shift by $a_0/2$ in the $z$ direction between two adjacent columns of atoms (see the left part of Figure~\ref{fig:Geometry}). Starting from a DFT calculation, Chulkov~\textit{et~al.}~\cite{des:quant:chulkov:99} have obtained a model potential in the $z$ direction where the periodicity is $a_0/2$. With this potential in the $z$ direction, the tridimensional potential $V(u,v,z)$, now separable in the first approximation, is used in the present simulation.

Secondly, solving the time-independent Schr\"odinger equation in this potential yields the set of wave functions defined in Eqs.~(\ref{eq:basis_functions}) and~(\ref{eq:f_I}). If one discretizes the bands of the solid using a DVR approach~\cite{dvr:bacic:light:1986,dvr:colbert:miller:1992}, the Chulkov potential for the e-Cu(001) system accommodates discretized bulk and image states. In the $z$ direction our grid extends from 70\,\AA\, in the vacuum to the 200\,\AA\, in the bulk, with 5881 grid points. As the initial state we select the last occupied discretized bulk state below the Fermi energy located at -0.12~eV from the Fermi energy already designated under the name of bulk Fermi state. The final state is the $n=1$ image state located at -0.574~eV from the vacuum energy. These two states are separated by 4.169~eV and their densities are presented in the lower part of figure \ref{fig:total_density_cu_001}.

In the direction parallel to the surface we use a periodic wave function given in Eq.~(\ref{eq:f_I}). We do not perform an explicit calculation to obtain the optimized wave functions. Instead, in the elementary cell, the localized wave functions $\chi_I(u,v)$ are two-dimensional harmonic oscillator wavefunctions. Following the usual selection rules, we have restricted this basis set to the $s$ and $p$ symmetries only
\begin{subequations}
\begin{eqnarray}
\label{eq:gaussian}
\chi_s(u,v)     & = & \sqrt{\frac{2\xi}{\pi}}\;e^{-\xi(u^2+v^2)}\\
\chi_{p_u}(u,v) & = & \sqrt{\frac{8}{\pi}}\,\xi\;u\;e^{-\xi(u^2+v^2)}\\
\chi_{p_v}(u,v) & = & \sqrt{\frac{8}{\pi}}\,\xi\;v\;e^{-\xi(u^2+v^2)}
\end{eqnarray}
\end{subequations}
As shown in Eq.~(\ref{eq:f_I}), the periodic wave function $f_I(\vec{r}_{\parallel})$ is obtained from these localized basis functions translating them by a multiple of $a_0^u$ and $a_0^v$ (see the caption of Figure~\ref{fig:Geometry}). On the Cu(001) surface this distance is $a_0^u=a_0^v=2.556\,$\AA\, and the symmetry of the surface cut imposes the use of the same exponent $\xi$ in the $u$ and $v$ directions. This exponent was chosen such that the resulting wave functions are confined in the elementary cell. For simplicity, and because no optimization is performed parallel to the surface, the energies $E_I^{\parallel}$ of the $s$, $p_u$ and $p_v$ states in Eq. (\ref{eq:diagonal}) are set to zero.

The vector potential $\vec{A}(\vec{r},t)$ has been obtained using the EMFED model following the prescriptions given in section \ref{sec:EMFED_model} where the electron density is calculated using the Eq.~(\ref{eq:electron_density}) and the wave functions $\eta_I(z)$ (Eq. (\ref{eq:basis_functions}) are obtained using the DVR calculations performed using the grid given above. At the photon excitation energy of 4.169~eV, the complex refraction index $\tilde{n}= 1.386+1.687\,i$ of the solid is taken from the tables of Palik~\cite{book:palik_opt_cont}. It is used to calculate the dielectric constant $\tilde{\epsilon}_s(\omega)$ of the solid and, combined with the electron density from Eq. (\ref{eq:electron_density}), the dielectric function $\tilde{\epsilon}(\omega,z)$ in Eq. (\ref{eq:dielectric_analytique}). Finally, the refraction index $\tilde{n}(\omega,z)$ is calculated from Eq. (\ref{eq:refraction_index}) and used to construct the EFMED vector potential in subsection \ref{sec:A_pot}. 

In the calculations, the incident angle of the laser beam is fixed at $\theta_i$=45 degrees except for the one presented in figure \ref{fig:observables_theta} where this angle is varied. For the Cu(001) surface, the observables calculated in the present work are independent of the orientation angle $\varphi$ of the POI (see Fig. \ref{fig:Geometry}). We have therefore fixed the orientation of the plane of incidence (POI) parallel to the $u$ axis of the material system i.e. $\varphi$=0. With these parameters, the vector potential $\vec{A}(\vec{r},t)$ is calculated using Eqs.~(\ref{eq:A_z}-\ref{eq:A_y}). We consider a pulsed laser with a peak intensity of 10$^8$~W/cm$^2$ and a temporal FWHM of $\tau=80$~fs resulting in a fluence of 10~$\mu$J/cm$^2$. This fluence is well within the perturbation regime, where space charges near the electron analyzer do not influence TPPE measurements~\cite{velic:wolf:98,kirchmann:wolf:2005}.

The calculations are performed at the $\Gamma$ point and the dispersion is neglected. Therefore, in equation (\ref{eq:integral_7}) of the appendix, only the first term is calculated. With the present choice of basis functions parallel to the surface, the laser-matter interaction terms between the $s$ and $p_u$ or $p_v$ states, given in Eq.~(\ref{eq:integral_8}), reduce to
\begin{equation}
{^u}\mathcal{D}^{00,00}_{IJ} = \int\!\!\!\!\int \chi^*_{s}\;\frac{\partial}{\partial u}\;\chi_{p_u}\;du\,dv = \sqrt{\xi}\,,
\end{equation}

where $s$ and $p_u$ belong to different states $I$ perpendicular to the surface. The ODE system of coupled equations~(\ref{eq:coupled_equations}) contains six states, three $s$, $p_u$ and $p_v$ for the initial and the final states. In the unambiguous cases these initial and final states will be referred simply to as $i$ and $f$. The integrals over the basis functions for the e-surface potential and the laser-matter interaction have been obtained once for all and the system of six coupled ODE~(\ref{eq:coupled_equations}) has been solved at each time step by a computer program based on a predictor-corrector algorithm~\cite{ab:shampine:burkardt:1975}. The obtained time dependent expansion coefficients permit the calculation of the excitation probability (Eq.~(\ref{eq:section_efficace_I})) and of the current density (Eqs.~(\ref{eq:courant_I})-(\ref{eq:courant_I_int})).
\begin{figure}[htb!]
\centering
\includegraphics[width=8cm,clip]{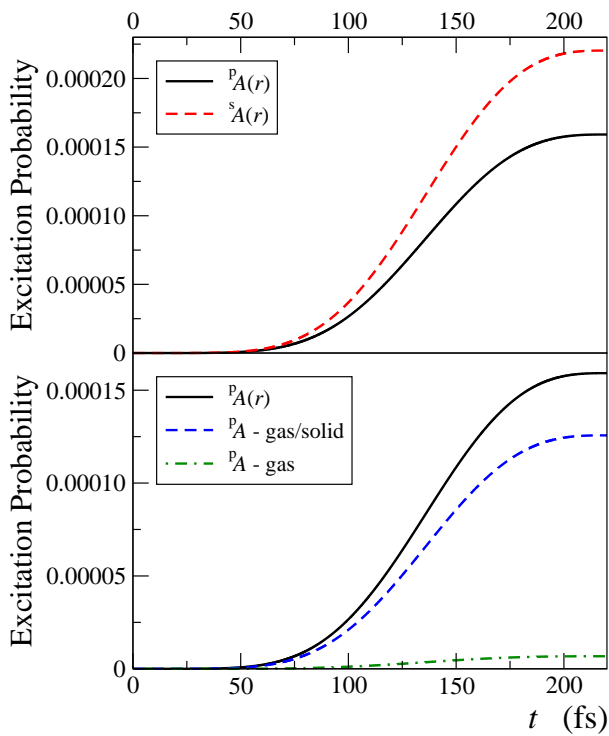}
\caption{(color online) Temporal evolution of the excitation probability at resonant energy $\hbar\omega= E_{f}-E_{i}=4.169$~eV for a pulsed laser FWHM of $\tau=80$~fs and intensity of \mbox{10$^8$ W/cm$^2$}. The displayed time interval corresponds to a total duration of 220~fs. Upper graph: Excitation probability calculated with the EMFED vector potential with the laser polarizations $\mathfrak{p}$ and $\mathfrak{s}$. The direction of the incident light is $\theta_i$=45 degrees and the POI is parallel to the $(100)$ direction that corresponds to $\varphi=0$ in Figure~\ref{fig:Geometry}. Lower graph: Excitation probability calculated with the $\mathfrak{p}$-polarization for the EMFED vector potential (black full line), for an abrupt vector potential at the interface (gas/solid, blue dashed line) and for a constant vector potential from the gas phase (gas, green dotted dashed line).}
\label{fig:res_excitation_probab}
\end{figure}

\subsection{Excitation dynamics}
%
\begin{figure}[htb!]
\centering
\includegraphics[width=8cm,clip]{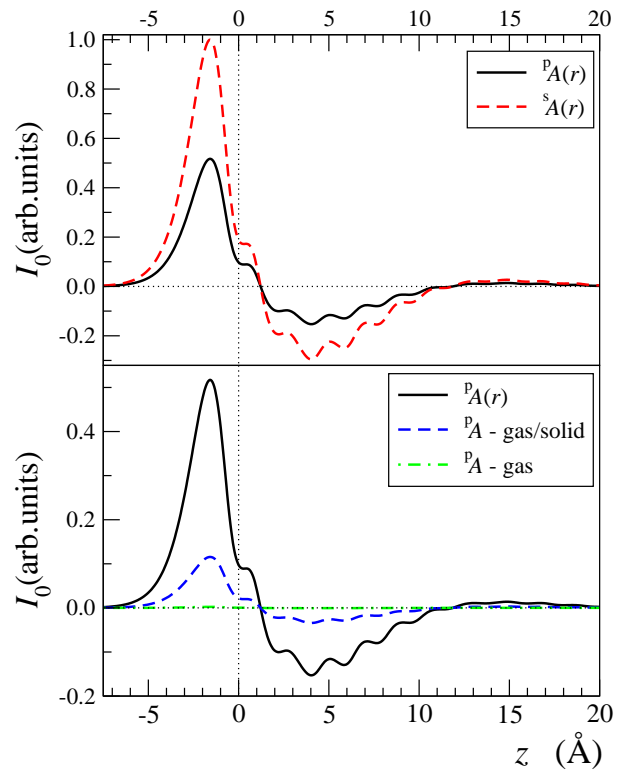}
\caption{(color online) Current density $I_0$, corresponding to the kinetic energy operator $\nabla^2$ obtained from the Eq.~(\ref{eq:courant_I_0}), as a function of the coordinate $z$ at the peak of the laser pulse. This current density is normalized to the maximum of $^\mathfrak{s}\!I_0$ using the normalization factor of 6.5$\times$10$^{-5}$.  The characteristics of the laser are given in the caption of Figure~\ref{fig:res_excitation_probab}.}
\label{fig:res_current_0}
\end{figure}
\begin{figure}[htb!]
\centering
\includegraphics[width=8cm,clip]{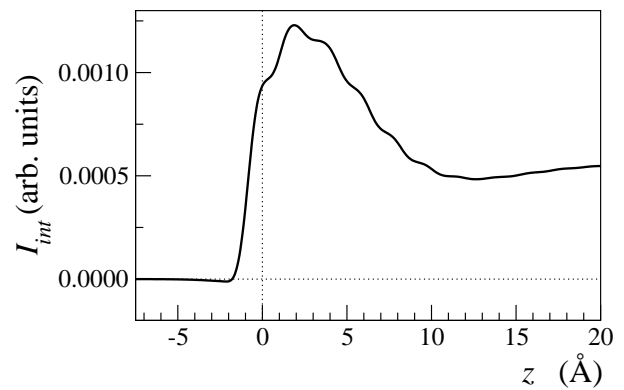}
\caption{Interaction current density $I_{int}$ for the $\mathfrak{p}$ polarization of the laser corresponding to the operator $2\;\vec{A}\cdot\vec{\nabla} + (\vec{\nabla}\cdot\vec{A})$ at the peak of the laser pulse as a function of the coordinate $z$ calculated from the Eq. (\ref{eq:courant_I_int}). This current density is normalized using the factor 3.37$\times$10$^{-5}$. The characteristics of the laser pulse are given in the caption of Figure~\ref{fig:res_excitation_probab}.}
\label{fig:res_current_int}
\end{figure}
\begin{figure}[htb!]
\centering
\includegraphics[width=8cm,clip]{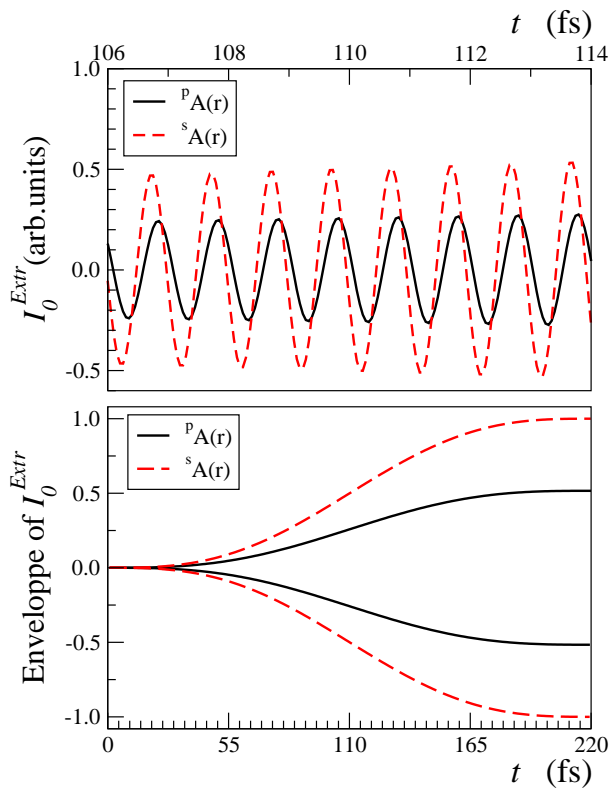}
\caption{(color online) Temporal evolution of the extrema of the current density $I_0$ at resonance $\hbar\omega=4.169$~eV for the $\mathfrak{p}$- and $\mathfrak{s}$- polarizations of the laser. The characteristics of the laser pulse are given in the caption of Figure~\ref{fig:res_excitation_probab}. The current density is normalized to the maximum of $^\mathfrak{s}\!I_{0}$ using the normalization factor 1.31$\times$10$^{-4}$. Upper graph: the displayed temporal region corresponds to the peak of the laser pulse. Lower graph: envelope of the temporal evolution of the extrema of the current density $I_0$.}
\label{fig:res_current_0_temp}
\end{figure}

The excitation probability between the bulk Fermi state and the first image state has been calculated at resonance for the two laser polarizations $\mathfrak{p}$ and $\mathfrak{s}$. The evolution of this excitation probability during the laser pulse is presented in Figure~\ref{fig:res_excitation_probab}. It is the sum of the excitation probabilities of the $s$, $p_u$ and $p_v$ image states calculated using the Eq.~(\ref{eq:section_efficace_I}). This observable does not display the oscillatory pattern related to the phase of the laser field since the square in eq. (\ref{eq:section_efficace_I}) annihilate the phase of the expansion coefficients and the rise and the stabilization of the excitation probability is smooth.

As seen in the upper part of Figure~\ref{fig:res_excitation_probab}, the excitation probabilities of the $\mathfrak{s}$ and $\mathfrak{p}$- polarizations are different. The expression of the excitation probability for the $\mathfrak{s}$ polarization contains a single contribution from the $A_y$ component of the vector potential. For $\mathfrak{p}$ polarization both $A_x$ and $A_z$ contribute, giving rise to a destructive interference which explains the present lower probability for this polarization.

In the lower part of Figure~\ref{fig:res_excitation_probab}, one compares the excitation probability for the $\mathfrak{p}$ polarization calculated with the vector potential of the EMFED model, with the one calculated with a discontinuous vector potential at the interface (label gas/solid, see also Fig.~\ref{fig:cu_001_As_y}). A comparison is also performed with a vector potential independent of the spatial coordinate (label gas). The excitation probability calculated using the gas/solid discontinuous potential gives a result 25 \% lower than that calculated with EMFED. The simulation using the spatially independent constant vector potential gives an excitation probability 24 times smaller than that obtained using the EMFED model. Taking into account the spatial variation of the laser field at the interface is therefore crucial for a realistic description of the electron dynamics at the interface.

Moreover we have calculated the current density decomposed into two contributions: $I_0$ and $I_{int}$. The current density $I_0$ is three orders of magnitude larger than $I_{int}$ and therefore $I_0$ presents all the characteristics of the total current density. The spatial dependence of the current density $I_0$ and $I_{int}$ at the peak of the laser pulse, obtained from the expressions (\ref{eq:courant_I_0}) and (\ref{eq:courant_I_int})), is presented respectively in Fig.~\ref{fig:res_current_0} and \ref{fig:res_current_int}. 

The $I_0$ current density displays spatial oscillations in the $z$ direction that mimic the oscillations in the electron-solid interaction potential shown in the lower part of Fig.~\ref{fig:total_density_cu_001}. The $I_{int}$ current density shows similar but attenuated oscillations. The Eqs. (\ref{eq:courant_I_0}) and (\ref{eq:courant_I_int}) show that the $I_0$ and $I_{int}$ current densities are calculated as definite integrals from $-\infty$ to $z$. Changing the upper limit $z$ adds positive or negative contributions to the current giving rise to the observed oscillations. As for the excitation probability in Figure~\ref{fig:res_excitation_probab} and for the same reasons the $\mathfrak{s}$ current density is larger than the $\mathfrak{p}$ current density. Outside the surface region the current density is zero therefore at a nanoscopic scale there is no charge migration.

The lower part of Fig.~\ref{fig:res_current_0} displays the $I_0$ current density for the different approximations of the vector potential. One sees a similar behavior to the one obtained for the excitation probability. These two observables are of different nature and one concludes that the introduction of the variation in space of the vector potential is essential for a proper modeling at the gas-solid interface.

Figure~\ref{fig:res_current_int} presents the current density $I_{int}$ for the $\mathfrak{p}$ polarization of the laser. For the $\mathfrak{s}$ polarization the corresponding current density $I_{int}$ is zero and is not displayed. This result can be explained as follows. For the $\mathfrak{p}$- polarization, the numerical derivatives of the basis functions, relative to the $z$ coordinate, do not compensate in Eq.~(\ref{eq:courant_I_int}) resulting in a non-zero interaction current density $I_{int}$. The zero contribution for the $\mathfrak{s}$- polarization is traced back to the integral Eq.~(\ref{eq:courant_I_int}) where in this case the two terms compensate each other due to the conservation of the momentum parallel to the surface (see appendix).
\begin{figure}[htb!]
\centering
\includegraphics[width=8cm,clip]{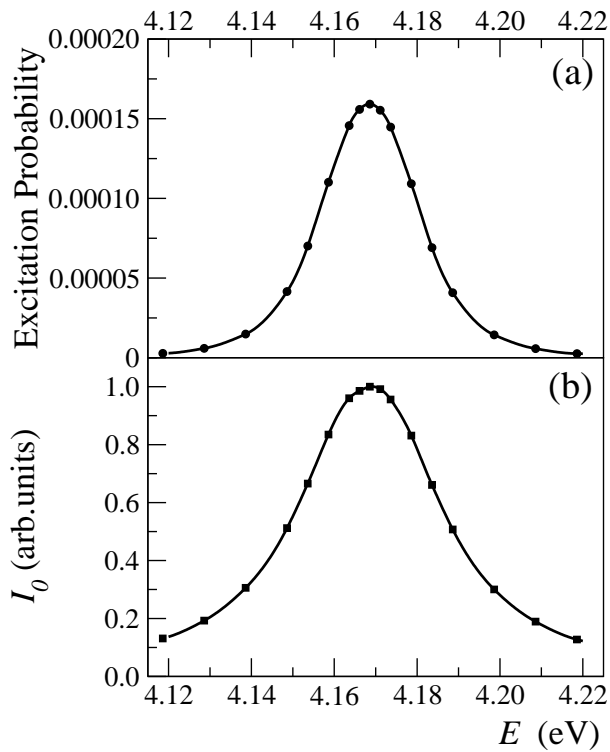}
\caption{(color online) Energy behavior near the resonance energy at \mbox{$\hbar\omega=4.169$~eV} of the excitation probability for $t\rightarrow\infty$ (a) and of the maximum of the current density $I_0$ (b) approximately located in the vacuum at $z\simeq$-2~\AA.  This current density is normalized to the maximum of $^\mathfrak{p}\!I_0$ using the normalization factor 6.8$\times$10$^{-5}$. The characteristics of the laser pulse are given in the caption of Fig.~\ref{fig:res_excitation_probab}.}
\label{fig:energy_var_observables}
\end{figure}
\begin{figure}[htb!]
\centering
\includegraphics[width=8cm,clip]{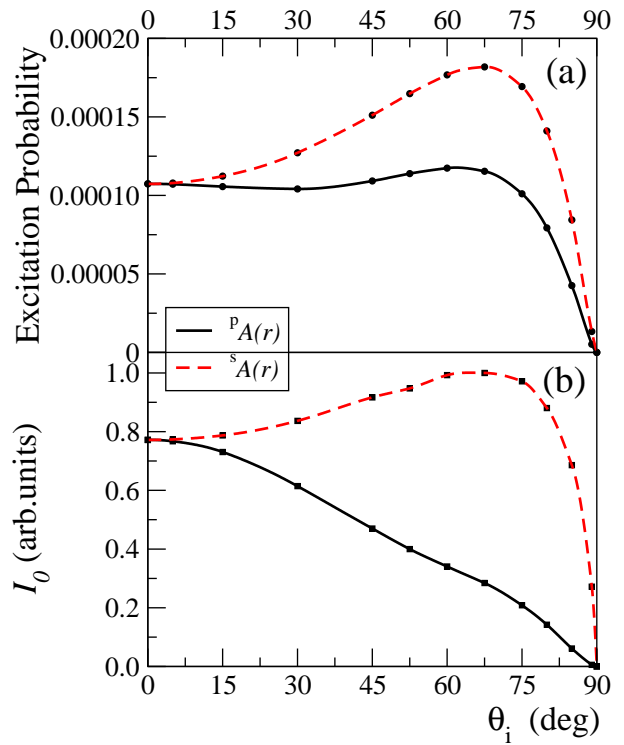}
\caption{Excitation probability for $t\rightarrow\infty$ (a) and maximum of the current density $I_0$ (b) located in the vacuum approximately at $z\simeq$-2~\AA\, and normalized with the factor 1.2$\times$10$^{-4}$ as a function of the incident angle $\theta_i$ of the laser beam. The calculations are performed at the excitation energy \mbox{$\hbar\omega=4.179$~eV} and the characteristics of the laser pulse are given in the caption of Figure~\ref{fig:res_excitation_probab}. The calculations correspond to the laser polarizations $\mathfrak{p}$ (full black line) and  $\mathfrak{s}$ (dashed red line).}
\label{fig:observables_theta}
\end{figure}

Next we present in Figure~\ref{fig:res_current_0_temp} the temporal evolution of the extrema of $I_0$, around $z\simeq$ -2~\AA. The upper graph presents a small region of the temporal evolution of these extrema near the peak of the pulse for the $\mathfrak{p}$- and $\mathfrak{s}$- polarizations. As expected, the current density oscillates at the Bohr frequency $\omega_B= (E_f-E_i)/\hbar\simeq$~6.3 fs$^{-1}$. The calculations for the different polarizations appear in the same order as the excitation probability. Moreover, there is a small temporal dephasing between the $\mathfrak{p}$- and $\mathfrak{s}$- polarizations. A possible origin for this time shift is the destructive interference between the $^{\mathfrak{p}}\!A_z$ and $^{\mathfrak{p}}\!A_x$ contributions to the current (see Eqs.~(\ref{eq:interaction}), (\ref{eq:J_int_p}) and (\ref{eq:J_int_s})). In opposition to the excitation probability where the laser phase is lost, the observed oscillatory behavior in $I_0$ is due to the relative phase of the products of the complex expansion coefficients $C_i^*\;C_f$ or $C_f^*\;C_i$ in Eq.~(\ref{eq:courant_I_0}). The envelope of the evolution of these extrema of the current density $I_0$, presented in the lower part of Figure~\ref{fig:res_current_0_temp}, displays a smooth variation similar to the evolution of the excitation probability (Figure~\ref{fig:res_excitation_probab}). Because the interaction current $I_{int}$ is about three order of magnitude smaller than $I_0$ (see Fig.~\ref{fig:res_current_0} and~\ref{fig:res_current_int}), we do not present the similar temporal variation of this current.

A final series of results is shown in Figures~\ref{fig:energy_var_observables} and~\ref{fig:observables_theta}. The first figure displays the behavior of the extrema of the calculated observables as a function of the photon energy near the resonant excitation energy between the bulk Fermi state and the $n=1$ image state. The excitation probability at t$\rightarrow\infty$ and the maximum of the current density $I_0$ display the usual behavior where a maximum appears at the resonance energy. The FWHM of these Lorentzian-like curves for the excitation probability and for the current density are 0.026~eV and 0.035~eV, in agreement with the bandwith of the pulse.

The second figure displays the evolution of the excitation probability and of the maximum current density $I_0$ with the incident angle $\theta_i$ of the laser beam (see Fig.~\ref{fig:Geometry}). As one expects, when the incidence is normal to the surface, the observables have the same value for the $\mathfrak{p}$- and $\mathfrak{s}$- polarizations. At grazing incidence the laser field does not penetrate in the solid and no excitation occurs. But as soon as the incident angle $\theta_i$ diminishes, for example for  $\theta_i=89\,$deg these observables rise sharply. 

\section{Discussion and conclusion}

This paper presents a detailed analysis of the laser-matter interaction at the gas-solid interface in the long wavelength domain, i.e. $\lambda \geqslant 100$~\AA. The main goal of the present work is the identification of the role played by the spatial variation of the laser fields, due to the sudden rise of the electron density at the surface, on the electron dynamics at the gas-solid interface.

First let us briefly recall the model. For a one photon excitation, our theoretical model calculates the excitation probability and the associated electron current density between a state of the solid near the Fermi energy (called bulk Fermi state) and the $n=1$ image state. The laser-matter interaction Hamiltonian (subsection~\ref{sec:hamiltonian}) is calculated using the vector potential of the laser field obtained from the EMFED model as a function of the $z$ coordinate (section~\ref{sec:EMFED_model}). The electron dynamics at the interface is obtained by the solution of the time dependent Schr\"odinger equation  (subsection~\ref{subsec:time_dependent_model}).

For the Cu(001) surface, the present model makes use of a separable potential parallel and perpendicular to the surface (see subsection \ref{subsec:physmod}), the last component of the potential being taken from the work of Chulkov~\textit{et~al.}~\cite{des:quant:chulkov:99}.

From the total unperturbed electron density $\rho_e(\vec{r})$ of the material system, the EMFED model obtains, using the constants of the material, a realistic vector potential $\vec{A}(\vec{r},t)$ function of the spatial coordinate for an arbitrary incident angle of the photon. The only needed ingredient is an electron-matter interaction potential.

The total wave function is expanded (Eq.~(\ref{eq:expansion})) in a series of six discrete basis functions (see subsections~\ref{subsec:time_dependent_model} and \ref{subsec:physmod}). The first three are constructed as a simple product of the ground bulk Fermi state in the perpendicular direction and a $s$ or $p$ orbital parallel to the surface. The last three are constructed similarly but with the $n=1$ image state in the perpendicular direction. The projection on these basis functions of the time dependent Schr\"odinger equation gives rise to a system of coupled first order ordinary differential equations in time (Eq.~(\ref{eq:coupled_equations})) which are solved for each time step by a predictor corrector algorithm.

Next, the results of the present model can be summarized as follows. The analysis of Figures~\ref{fig:res_excitation_probab}, \ref{fig:res_current_0} and~\ref{fig:res_current_int} concludes that, near the gas-solid interface, a realistic model of the electron dynamics in the presence of the laser-matter interaction requires the inclusion of the spatial variation of the laser field even in the long wavelength domain.

The excitation probability (Fig.~\ref{fig:res_excitation_probab}) and the current density (Fig.~\ref{fig:res_current_0}) are different for the $\mathfrak{p}$ and $\mathfrak{s}$- polarizations of the laser, giving an indication on the so called surface photoelectric effect appearing in the neighborhood, approximately $\pm$~3--5~\AA, of the interface. Precisely, Table \ref{tab:A_p_operator} gives the different contributions to the interaction Hamiltonian and one sees that, even if the vector potential projections $x$ and $y$ are equivalent, the $x$ scalar product contains the derivative with respect to $x$ of the vector potential whereas the $y$ scalar product does not contain a similar surface term. Therefore, from the difference between the observables for the $\mathfrak{p}$ and $\mathfrak{s}$- polarizations, in the present excitation between discrete states or an excitation giving rise to an electron emission in the vacuum, one cannot estimate precisely the surface photoelectric effect.

The current density $I_0(z,t)$ (Fig.~\ref{fig:res_current_0}) reproduces the spatial oscillations of the electron-solid potential as well as the temporal oscillations of the laser field. Contrary to the excitation probability, this observable is unravel the phase of the time evolution of the laser. The current density $I_{int}$ (Fig.~\ref{fig:res_current_int}) is directly related to the laser-matter interaction term in the Hamiltonian and, for photoemission, to the surface photoelectric effect. But this term is weak and cannot be measured experimentally.

Finally, let us discuss the approximations  and the limitations of the present model. First, the separable electron-surface potential contains a contribution in the $z$ direction explicitly calculated and a contribution parallel to the surface that is not explicitly calculated. But this separable potential could be replaced by a more realistic tridimensional potential obtained using, for example, a DFT calculation. In addition, the present simulation of the excitation between discrete states can easily be extended to the photoemission and, because our wave function is expanded on a basis set, the number of these basis functions can easily be raised.

Secondly, the use of tabulated material constants in our model implies that this model is able to account for the presence of the plasma oscillations in the bulk as one of us has shown for a Jellium Al(100) surface~\cite{emed:al:raseev:2005}. One can also account for the change in the observables at the surface plasmon frequency because in the present model the dielectric function depends on $z$. The dependence introduced here is similar to the one used by Feibelman~\cite{feibelman:1989} when he derives the frequency of the surface plasmon.

Let us now replace the present work in a more general context. The consequences of the spatial variation of the vector potential have been highlighted through a numerical calculation of the excitation probability and of the current density. But other physical observables \emph{qualitatively} sensitive to such a spatial variation of the vector potential, have to be identified, calculated and measured. For example, time resolved TPPE experiments, coupled with some interference measurements~\cite{pontius:petek:2005} or combination of TPPE and photoemission electron microscope (PEEM)~\cite{Munzinger:Aeschlimann:Bauer:2005, rohmer:aeschlimann:bauer:2006}, could be efficient experimental techniques for observables sensitive to the spatial variation of the vector potential. Particularly the last cited technique TPPE/PEEM allows for an imaging of single nano objects of a size of about 50~nm thus diminishing the averaging made over many objects in standard experiments. Combination of a laser with a STM~\cite{stm:gerstner:00,riedel:dujardin:05} in a single simultaneous experiment can be also used for measuring the observables related to these nano objects.

One can easily extend the present model to single photon photoemission but such a model remains oversimplified for the previously cited experimental techniques. In fact, to model a TPPE experiment, one has to take into account, in addition to the direct excitation processes and associated continua of the electron, the electron-electron relaxation at the surface and in the bulk. Modeling such an experiment, implies the combination of the EMFED vector potential with a density matrix formulation~\cite{wolf:1999,klamroth:saalfrank:hofer:2001,boger:weinelt:2002,pontius:petek:2005,boger:fauster:weinelt:2005,mii:ueba:2005}. Our approach calculates some of the ingredients needed for such a density matrix model, as for example the unperturbed states and the interaction matrix elements with a spatially varying vector potential. Other needed ingredients, like the elastic and inelastic parameters of the electron-electron collisions corresponding to the relaxation to the bath, can be taken from the existing literature.

Concerning other systems to be studied, different clean metallic surfaces can be selected, but adsorbates~\cite{berthold:hofer:2004,kirchmann:wolf:2005} or nano objects on surfaces~\cite{kubo:petek:2005,Munzinger:Aeschlimann:Bauer:2005,rohmer:aeschlimann:bauer:2006} could also be very sensitive to a spatially varying vector potential. The calculation of the observables as a function of photon frequency, particularly at the bulk and surface plasmon resonance, is the next goal in our modeling. Generalization to semiconductor surfaces is also of interest. However, for these surfaces reconstruction takes place and the wave functions parallel to the surface should be functions of the slab layer.

In summary, the present model clearly shows that the spatial variation of the vector potential, in the vicinity of the gas-solid interface, significantly modifies the electron dynamics near the surface, even in the long wavelength domain. Here, we have proposed a relatively simple approach for the calculation of this spatial variation. Future developments will allow for a better estimate of its influence on the measurements performed with recent experimental techniques like two-photon photoemission (TPPE).

\begin{acknowledgments}
We thank Herve le Rouzo for many fruitfull discussions. Discussions with Doina Bejan during the early stage of this work are also warmly acknowledged. This work has been done with the financial support of the LRC of the CEA, under contract number DSM 05--33. 
\end{acknowledgments}

\appendix
\section{Evaluation of the integral $\mathcal{J}_{IJ}(z)$}
\label{annex:A}

To obtain $\mathcal{J}_{IJ}(z)$, let us rewrite the vector potential $\vec{A}(x,z,t)$ of Eqs. (\ref{eq:A_z})-(\ref{eq:A_y}) in section~\ref{sec:A_pot} as a product of two terms related to the variables $z$, $x$ and $t$
\begin{equation}
\label{A_z_x_t}
^{\mathfrak{pol}}\!\vec{A}(x,z,t) = \;^{\mathfrak{pol}}\!\!\vec{\mathcal{A}}(z)\;\;
			e^{i\left(k_x^{ph}\,x\,-\,\omega\,t\right)}\,,
\end{equation}
Then, the interaction Hamiltonian $\hat{H}^L$ (Eq.~(\ref{eq:H_L})) becomes
\begin{equation}
\hat{H}^L = \frac{\hbar\;e}{2\;m\;i}\;\;^{\mathfrak{pol}}\!\mathcal{O}(\mathcal{A})\;\;
			e^{i\left(k_x^{ph}\,x\,-\,\omega\,t\right)}
\end{equation}
where
\begin{subequations}
\begin{eqnarray}
^{\mathfrak{p}}\!\mathcal{O}(\mathcal{A}) &=& 
			   \,^{\mathfrak{p}}\!\!\mathcal{A}_x\;
				\biggl[2\cos\varphi\frac{\partial}{\partial u}
			+ 	2\sin\varphi\frac{\partial}{\partial v} + i\,k_x^{ph}\biggr]
\nonumber \\
			&& + 2\,^{\mathfrak{p}}\!\!\mathcal{A}_z\frac{\partial}{\partial z}
				+ \frac{\partial\,^{\mathfrak{p}}\!\!\mathcal{A}_z}{\partial z},
\end{eqnarray}
and
\begin{equation}
^{\mathfrak{s}}\!\mathcal{O}(\mathcal{A}) = 2\;^{\mathfrak{s}}\!\mathcal{A}_y\;
				\biggl[-\sin\varphi \frac{\partial}{\partial u} 
			+ 	\cos\varphi \frac{\partial}{\partial v}\biggr],
\end{equation}
\end{subequations}
respectively for the $\mathfrak{p}$- and $\mathfrak{s}$- polarizations. In the equations above, the operator $^{\mathfrak{pol}}\!\mathcal{O}(\mathcal{A})$ is written in the orthogonal crystallographic coordinate system \{u,v,z\} of the surface (see Fig. \ref{fig:Geometry}), rotated by an angle $\varphi$ relative to the POI coordinate system. The angle $\varphi$ between the two coordinate systems can be varied in an experimental set-up. For the discrete transitions presented in this paper, the observables, transition probability and current density, are in principle independent of the angle $\varphi$. In photoemission, one can unravel the local  surface symmetry (see for example Smith {\it et al.} \cite{smith:76}).

The integral $\mathcal{J}_{IJ}(z)$, appearing in the laser-matter interaction matrix elements~(\ref{eq:interaction}), can now be written in the particular case of $\mathfrak{p}$- and $\mathfrak{s}$- polarizations as 
\begin{subequations}
\begin{eqnarray}
\label{eq:J_int_p}
^{\mathfrak{p}}\!\mathcal{J}_{IJ}(z) & = & \bigl(\, ^{\mathfrak{p}}S_{IJ}^{\perp}(z)\,^{\mathfrak{p}}\mathcal{D}_{IJ}^{\parallel}
+\,^{\mathfrak{p}}\mathcal{D}_{IJ}^{\perp}(z)\;	^{\mathfrak{p}}\!S_{IJ}^{\parallel}\bigr)\mathcal{T}^{\parallel}_{IJ}~~~~~~\\
\label{eq:J_int_s}
^{\mathfrak{s}}\!\mathcal{J}_{IJ}(z) & = & ^{\mathfrak{s}}S_{IJ}^{\perp}(z)\,^{\mathfrak{s}}\mathcal{D}_{IJ}^{\parallel}
\mathcal{T}^{\parallel}_{IJ}.
\end{eqnarray}
\end{subequations}
Taking into account the simplifications related to the LWL approximation, one can evaluate these integrals along the following lines:
The factor $\mathcal{T}^{\parallel}_{IJ}$ gives a selection rule for the angular momenta parallel to the surface
\begin{eqnarray}
\label{T_factor}
\mathcal{T}^{\parallel}_{IJ} &=& \sum_{j=0}^{\infty}\exp[i(\vec{k}_{\parallel}^I\;-\;\vec{k}_{\parallel}^J\;
		+\;\vec{k}_{x}^{ph})\cdot\vec{R}_{\parallel}^j]
\nonumber \\
	&\simeq&\delta(\vec{k}_{\parallel}^I-\vec{k}_{\parallel}^J+\vec{k}_{x}^{ph}).
\end{eqnarray}
The $\vec{k}_{x}^{ph}$ contribution appears as a consequence of the presence of exponential $\exp(i\;k_x^{ph}\;x)$ in the spatially varying vector potential Eq. (\ref{A_z_x_t}). In the LWL approximation this exponential is a slowly varying quantity relative to the surface reciprocal vectors  $k_u$ and $k_v$ and one can use the mean value theorem to factorize it from the integrals $^{\mathfrak{pol}}\!\mathcal{D}_{IJ}^{\parallel}$ or $^{\mathfrak{pol}}\!S_{IJ}^{\parallel}$. However, because $k_x^{ph}\ll  k_{\parallel}$, this condition reduces to the standard condition of the conservation of the linear momentum  $\vec{k}_{\parallel}^I \simeq \vec{k}_{\parallel}^J$ parallel to the surface.
For $\mathfrak{p}$- polarization, the explicit expressions of the integrals Eq. (\ref{eq:J_int_p}) are
\begin{eqnarray}
\label{eq:integral_1}
^{\mathfrak{p}}\!S_{IJ}^{\perp}(z)
	&=&	\int_{-\infty}^z\;\eta_I^*(z')\mathcal{A}_x(z')\eta_J(z')\;dz'
\\  \label{eq:integral_2}
^{\mathfrak{p}}\!\mathcal{D}_{IJ}^{\parallel}	&=&  2\,(
	\cos\varphi\; \mathcal{D}_{IJ}^u	+	\sin\varphi\;\mathcal{D}_{IJ}^v) 
\nonumber \\
	&&	+  i\;k_x^{ph}\;S_{IJ}
\\  \label{eq:integral_3}
^{\mathfrak{p}}\!\mathcal{D}_{IJ}^{\perp}(z)&=& 
	\int_{-\infty}^z\;
		\eta_I^*(z')\biggl[2\;\mathcal{A}_z(z')\frac{\partial}{\partial z'}
\nonumber \\
	&&+\frac{\partial\mathcal{A}_z(z')}{\partial z'}\biggr]\eta_J(z')\;dz'
\\  \label{eq:integral_4}
^{\mathfrak{p}}\!S_{IJ}^{\parallel}&=&  2\,S_{IJ}(\cos\varphi + \sin\varphi) 
\end{eqnarray}
where $\mathcal{D}_{IJ}^u$, $\mathcal{D}_{IJ}^v$ and $S_{IJ}$ are function of the reciprocal lattice vectors $k_u$ and $k_v$. In the equations (\ref{eq:integral_1}) and (\ref{eq:integral_3}) one identifies the usual interaction terms present in the velocity gauge where the vector potential modulates the associated integrals. There are also new terms,  called surface contributions, originating from the derivative of the vector potential, like the last term in the Eq.~(\ref{eq:integral_3}).
For $\mathfrak{s}$- polarization, the integrals of Eq. (\ref{eq:J_int_s}) read
\begin{eqnarray}
\label{eq:integral_5}
^{\mathfrak{s}}\!\mathcal{D}_{IJ}^{\parallel}&=& 
		2\,(-	\sin\varphi\;\mathcal{D}_{IJ}^u	+	\cos\varphi\;\mathcal{D}_{IJ}^v)
\\ \label{eq:integral_6}
^{\mathfrak{s}}\!S_{IJ}^{\perp}(z)&=& \;\int_{-\infty}^z\;
	 		\eta_I^*(z')\;\mathcal{A}_y(z')\eta_J(z')\;dz'
\end{eqnarray}
No surface is term present for this polarization.
If one retains only the interaction with the near neighbors, the integrals  $\mathcal{D}_{IJ}^u$,  $\mathcal{D}_{IJ}^v$ and $S_{IJ}$, appearing in Eqs.(\ref{eq:integral_2}),  (\ref{eq:integral_4}) and (\ref{eq:integral_5}), present a similar structure. For example, this structure can be unravelled by writing explicitly the integral $\mathcal{D}_{IJ}^u$
\begin{eqnarray}
\label{eq:integral_7}
\mathcal{D}_{IJ}^u &=& {^u}\mathcal{D}^{00,00}_{IJ}
 \\
	&&+ 2\cos(k_u\,a_0^u)({^u}\mathcal{D}^{00,10}_{IJ} +  {^u}\mathcal{D}^{10,00}_{IJ})
\nonumber \\
	&&+ 2\cos(k_v\,a_0^v)({^u}\mathcal{D}^{00,01}_{IJ} + {^u}\mathcal{D}^{01,00}_{IJ})
\nonumber
\end{eqnarray}
where the integral ${^u}\mathcal{D}^{\ell_u\ell_v,\ell'_u\ell'_v}$ reads
\begin{eqnarray}
\label{eq:integral_8}
{^u}\mathcal{D}^{\ell_u\ell_v,\ell'_u\ell'_v}_{IJ} & = & \int\!\!\!\!\int \chi^*_{I}(u-\ell_u a_0^u,v - \ell_v a_0^v)\\
& & \frac{\partial}{\partial u}\;\chi_{J}(u-\ell'_u a_0^u,v - \ell'_v a_0^v)\;du\,dv.
\nonumber
\end{eqnarray}
Here the integration domain has been extended over three elementary cells in each direction to calculate the interaction with the near neighbors.
The interaction operator $^{\mathfrak{pol}}\!\mathcal{O}(\mathcal{A})$ contains symmetric (derivative of the vector potential) and antisymmetric contributions relative to the inversion of the system of coordinates. This last contribution is predominant. Consequently, in Eq.~(\ref{eq:basis_functions}) the set of basis functions parallel to the surface should contain symmetric and antisymmetric basis functions. In the simplest case, the functions parallel to the surface should contain ``s" and ``p" like basis functions.

Finally note that the above simple derivation has been inspired by the one given in the appendix~F of  Desjonqu\`eres and Spanjaard~\cite{book:desjonqueres_spanjaard} where the vector potential is independent of the spatial coordinates.


\begin{thebibliography}{46}
\bibitem{Drude}
In the Drude model, presented in the book of Ashcroft and Mermin~\cite{book:ashcroft:76}, the conductivity does not depend on the laser frequency. In the section~13.3 of Born and Wolf~\cite{book:born_wolf}, the expressions~(\ref{eq:Esigma}) and (\ref{eq:conductivity}) are considered to come from the Drude model.
\end{thebibliography}
\end{document}